# Physics-informed Neural Networks (PINNs) for Wave Propagation and Full Waveform Inversions

Majid Rasht-Behesht[1], Christian Huber[1], Khemraj Shukla[2] and George Em Karniadakis[3]

[1]Department of Earth Environmental and Planetary Sciences, Brown University, Providence, RI 02906, USA

[2]Division of Applied Mathematics, Brown University, Providence, RI 02906, USA

[3]Division of Applied Mathematics and School of Engineering, Brown University, Providence, RI 02906, USA

## Abstract

We propose a new approach to the solution of the wave propagation and full waveform inversions (FWIs) based on a recent advance in deep learning called Physics-Informed Neural Networks (PINNs). While PINNs have been successfully applied to several forward and inverse problems in science and engineering, their performance for solving the wave equation and FWIs, where labeled training data sets are often limited to observed data at the surface (boundary), has not been explored. In this study, we present an algorithm for PINNs applied to the 2D acoustic wave equation and test the model with both forward wave propagation and FWIs case studies. These synthetic case studies are designed to explore the ability of PINNs to handle varying degrees of structural complexity using both teleseismic plane waves and seismic point sources. PINNs' meshless formalism allows for a flexible implementation of the wave equation and different types of boundary conditions. For instance, our models demonstrate that PINN automatically satisfies absorbing boundary conditions, a serious computational challenge for common wave propagation solvers. Furthermore, *a priori* knowledge of the subsurface structure can be seamlessly encoded in PINNs' formulation. We find that the current state-of-the-art PINNs provide good results for the forward model, even though spectral element or finite difference methods are more efficient and accurate. More importantly, our results demonstrate that PINNs yield excellent results for inversions on all cases considered and with limited computational complexity. Using PINNs as a geophysical inversion solver offers exciting perspectives, not only for the full waveform seismic inversions, but also when dealing with other geophysical datasets (e.g., magnetotellurics, gravity) as well as joint inversions because of its robust framework and simple implementation.

## Glossary

**Free-surface constraint:** Stress-free condition imposed at the top of the model to simulate the surface of the Earth.

**Boundary data**: Any data for the wavefield located at the boundaries of the computational domain. This can be either in the form of observed data (seismograms), or in the form of imposed conditions such as a free-surface or absorbing conditions.

**Early-time snapshots**: Data from the snapshots of the wavefield recorded at t=0 and t=∆t obtained here with a wave equation solver (SpecFem2D) that are used as inputs to the PINN to train for characteristics of the seismic source in space and time.

## Introduction

Seismic inversions are important tools for imaging heterogeneities in the earth's subsurface. However, the resolving power of seismic inversion techniques is strongly controlled by the type and extent of the data used from seismograms. Particularly, in scenarios where the dominant wavelength of the seismic wave is greater than the length-scale of the target heterogeneities, FWIs can provide significantly better results than, for instance, travel-time tomography (Rasht-Behesht et al., 2020; Fichtner and Trampert, 2011; Marquering et al., 1999; Cassidy, 1992). Regardless of their advantages, FWIs remain technically and computationally challenging. Adjoint methods (Bozdag et al., 2016; Tromp et al., 2008; Fichtner et al., 2006; Plessix, 2006) offer an efficient strategy to reduce computational cost for FWIs from N forward simulations per optimization step (N= number of model parameters) to two forward simulations. However, the derivation and implementation of adjoint methods for FWIs remains challenging and must be treated on a case-by-case basis for different systems (Bozdag et al., 2016; Cockett et al., 2015).

With the successful application of machine learning, and in particular deep learning techniques, and the concurrent explosion of recorded seismic data made available over the past decade, seismologists have begun to search for modern efficient techniques (Kong et al., 2019; Bergen et al., 2019) to tackle problems such as earthquake detection, automatic phase picking (Mousavi et al., 2020; Ross et al., 2018; Yoon et al., 2015) and seismic signal denoising (Zhu et al., 2019). A common theme thus far in the majority of these applications has been the abundance of labeled/unlabeled training data sets. As a result, areas such as seismic imaging applications where one is limited to spatially sparse data sets (seismometers), have not benefited as much from these methodologies. However, under conditions where a computationally inexpensive forward model of the wave propagation exists, such as when considering a 1D layered earth structure, it is possible to generate large synthetic data sets to train a deep neural network (DNN) that can be then used as an efficient forward model surrogate (Moseley et al., 2020; Araya-Polo et al., 2018). The intent of this approach is to make the solution to the forward problem more cost effective than the numerical solution of the wave equation and thus accelerate the procedure for inverse problems in seismology. The drawback of this technique is that it heavily relies on preexisting fast forward solvers for training purposes, but once trained, the neural network (NN) can be used as an independent machinery. Another approach, to accelerate the calculations of the derivatives in gradient-based techniques such as adjoint-methods is the use of reverse-mode automatic differentiation efficiently implemented in current deep learning libraries such as TensorFlow (Abadi et al., 2016) and PyTorch (Paszke et al., 2019) to solve the adjoint state FWIs (Zhu et al., 2021).

Raissi et al., 2019 have introduced a new class of DNNs known as Physics-Informed Neural Networks (PINNs) that provide an improved framework to overcome the obstacles pertaining to the training of DNNs caused by challenges in training data acquisition. PINNs take advantage of the governing physical laws behind the processes that generate the data, hence reducing greatly the reliance of the DNN on the labeled data during the training process. PINNs simply bridge the gap between data-scarcity and the data-intensive nature of DNNs and have been successfully applied to various problems in engineering and biology with applications including but not limited to heat transfer (Cai et al., 2021), solution of Navier-Stokes equations in fluid mechanics (Jin et al., 2021), high speed fluid flow (Mao et al., 2020) and solid mechanics (Haghighat et al., 2021). For a recent review of PINNs application see Karniadakis et al., 2021.

Haghighat et al., 2021 explored PINNs' application to linear elasticity and nonlinear plasticity and showed that they can be efficiently applied to inversion and surrogate modeling in solid mechanics. Shukla et al., 2020 & 2021 used PINNs for a nondestructive quantification of surface cracks and identification of microstructural properties of polycrystalline Nickle using ultrasound data. Moseley et al., 2020, used PINNs as a solver for the forward acoustic wave propagation, while Smith et al., 2020 and Waheed et al., 2021 applied PINNs to the Eikonal equation as a forward solver for first arrival-time prediction and travel time tomography, respectively. Song et al., 2020 solved the frequency-domain anisotropic acoustic wave equation with PINNs. Nevertheless, to our knowledge, we present the first FWI for seismological applications using PINNs. In this study, we focus on the development of acoustic FWI with PINNs and demonstrate its practical application to various synthetic case studies. The salient results of our study are:

a) In most applications of PINNs, authors have incorporated training data sets from within the computational domain from other solvers or experimental data, which greatly facilitates the training process. In contrast, with seismic inversions, this is generally not possible (records of the wavefield are generally limited spatially to the surface or boreholes). We show that this limitation does not prevent PINNs from performing efficient and accurate seismic inversions.
b) We discuss the implementation of various types of boundary data, including a stress-free constraint on the top surface of the physical domain and absorbing boundary layers necessary for the simulation of wave propagation in infinite media. We show that the implementation of these constraints can be seamlessly handled with PINNs.
c) We present specific normalization guidelines that are crucial to PINNs' convergence as applied to the wave equation.
d) We demonstrate how to handle multiple seismic sources to improve the illumination of complex structural heterogeneity at depth.

## Methods

**Acoustic Wave Propagation**
The propagation of acoustic waves in a 2D medium with negligible density variations and in the absence of body forces can be described in terms of the scalar wave potential $\phi$ as:

$$\alpha^2 \nabla^2 \phi + f = \frac{\partial^2 \phi}{\partial t^2}, \qquad (1)$$

where $\nabla^2 \equiv \frac{\partial^2}{\partial x^2} + \frac{\partial^2}{\partial z^2}$ is the Laplacian operator defined in the cartesian coordinate system, $f$ is the external surface forces and $\alpha$ characterizes the acoustic wavespeed that strictly depends on the material properties of the medium through the relation $\alpha = \sqrt{\frac{K}{\rho}}$ with $K$ and $\rho$ being the material bulk modulus and density, respectively. Without loss of generality, we set $f \equiv 0$ and instead, enforce external forces through a perturbation of the initial field acting at some early time. The displacement field is retrieved from the gradient of the wave potential i.e., $(u_x, u_z) = \nabla \phi$.

The forward simulation of the wave propagation involves solving equation 1 given a set of boundary constraints and two early-time snapshots of the wave propagation as well as a precise knowledge of the material properties (wavespeed $\alpha$) of the medium in the computational domain. Note that the first time-snapshot constrains the shape of the seismic source whereas the second enforces its propagation direction. In the inverse problem, $\alpha$ is either unknown or there exists some a priori information and it is evaluated spatially from the recorded ground motion. In this study we will investigate how PINNs can be used to solve these two classes of problems.

**Neural Network (NN)**
PINNs' architecture has been almost exclusively based on fully connected feed-forward NN. In the absence of any justifiable reasons to do otherwise, we, therefore, define a fully connected feed-forward NN with an input layer consisting of the physical coordinates $x, z$ and time $t$, $L$ hidden layers and an output layer representing the scalar acoustic wave potential $\phi \in \mathbb{R}$ (Fig. 1). The various other physical variables, such as displacement or pressure, are obtained through the *automatic differentiation* of the wave potential NN using TensorFlow (Abadi et al., 2016). Note that the proper choice of the independent physical variable(s) that the output of the NN represents is problem-dependent. For instance, Haghighat et al., 2021 selected a separate NN for each component of the displacement field for problems in elasticity.

There are $N_l$ number of neurons in the $l^{th}$ hidden layer. A linear transformation followed by a nonlinear neuron-wise activation function ($\sigma$) is applied at the input $x^{l-1} \in \mathbb{R}^{N_{l-1}}$ to the $l^{th}$ layer:

$$H(x^{l-1}) = \sigma(w^l x^{l-1} + b^l),$$

where $w^l \in \mathbb{R}^{N_l \times N_{l-1}}$ is the matrix of weights and $b^l \in \mathbb{R}^{N_l}$ is the vector of biases corresponding to the $l^{th}$ layer. The successive operation of this transformation law results in the final output of the NN with a total number of $\sum_1^l (N_l \times N_{l-1}) + b^l$ tunable parameters ($N_0 = 3$). We choose $\sigma = \tanh(\cdot)$ or $\sin(\cdot)$ as the nonlinear activation function for all NNs used in this study. The activation function acting on the last hidden layer to yield the output is the identity function. The interested reader is directed to Jagtap et al., 2020a&b for a discussion of adaptive activation functions and their effect on convergence. The network's parameters are initialized from *independent and identically distributed (iid)* samples.

**PINNs for the forward problem**
We follow PINNs' original framework (Raissi et al., 2019), to obtain the parameters of a NN such that it closely approximates the acoustic wave potential $\phi$. To do so, we enforce this NN to satisfy the partial differential equation (1) with corresponding early-time snapshots of the wave propagation and boundary data computed on a set of randomly selected training data points (Fig. 2). In the following, we define the various residual terms that we aim to minimize:

$R_{PDE} \coloneqq \alpha^2 \nabla^2 \phi - \frac{\partial^2 \phi}{\partial t^2}$ 	PDE

$R_{P.C} \coloneqq \rho \alpha^2 \nabla^2 \phi(x, t, z = 0)$ 	Free-Surface Constraint

$R_{S_1} \coloneqq \nabla \phi(x, z, t = t_1^0) - \overrightarrow{U_1^0}(x, z)$ 	First time-snapshot

$$R_{S_2} := \nabla\phi(x,z,t=t_2^0) - \overrightarrow{U_2^0}(x,z) \qquad \text{Second time-snapshot}$$

$$R_{obs} := \nabla\phi(x,z,t) - \overrightarrow{U_{obs}}(x,z,t) \qquad \text{Observed data (For inversions)}$$

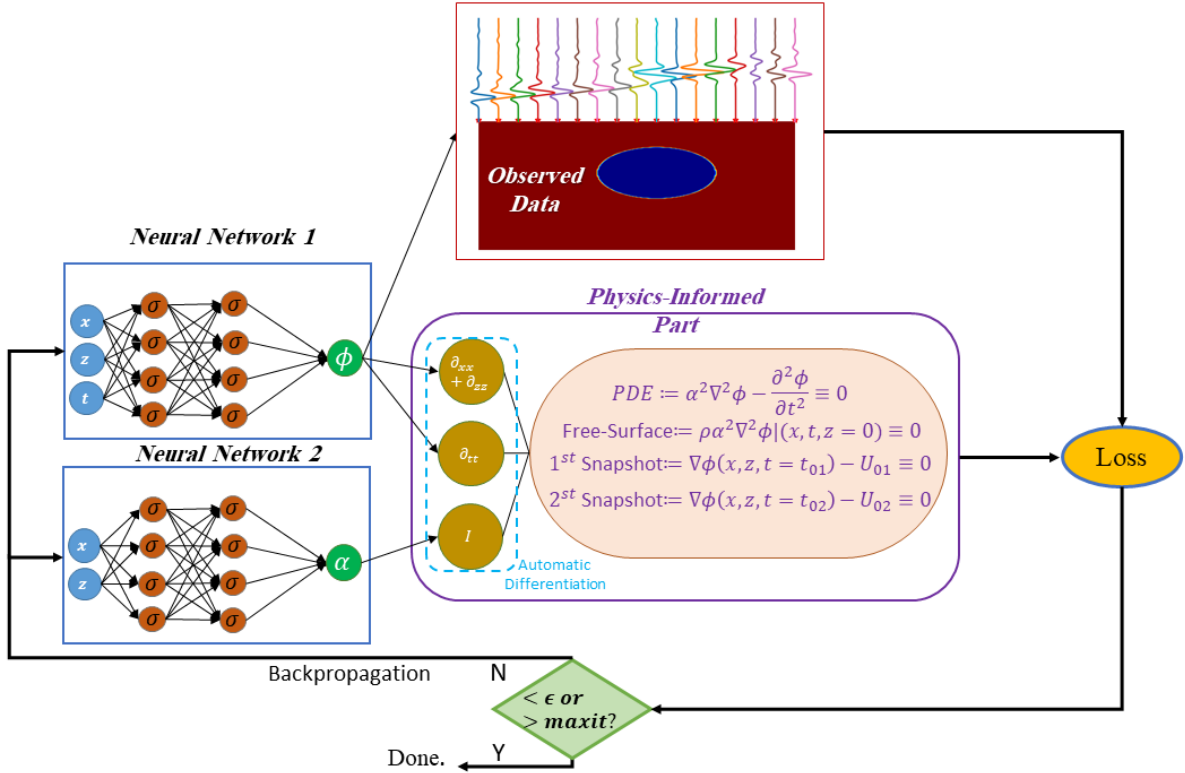

**Figure 1**. Schematics of PINNs' workflow. Left: Fully connected feed-forward neural network, the output of which approximates the solution to the forward and inverse problems. Right: The governing physical laws and the observed real-world data, i.e., seismograms, used to optimize the parameters of the PINN. The training stops when the loss error becomes smaller than a threshold, or the number of iterations goes beyond a set value.

$R_{P.C}$ indicates the physical free-surface constraint at the top of the model where pressure is fixed to zero. Observed data refers to the recorded wavefield on a sparse set of receivers (seismograms). In this study, the observed data in the form of synthetic seismograms and the early-time snapshots are obtained from SpecFem2D simulations (Komatitsch and Tromp, 1999; Tromp et al., 2008), a spectral element model for solving the wave equations in elastic/acoustic media. For the sake of convenience, we choose the time-snapshot data in terms of displacement, however one could choose one of these to explicitly involve the velocity field as well. Alternatively, one could also pose the training problem as an initial value problem, where the wavefield snapshots for training are provided in the form of two initial conditions at $t = 0$. It is important to note that modeling the wave propagation forward with only two early time snapshots of the wavefield cannot be implemented with traditional solvers, which illustrates the flexibility of PINNs in dealing with data at any time when applied to the wave equation.

The objective of the training process is to minimize the sum of mean squared errors:

$$MSE(\Theta) = \lambda_1 MSE_{PDE} + \lambda_2 MSE_S + \lambda_3 MSE_{P.C} + \lambda_4 MSE_{Obs}, \qquad (2)$$

Where $\Theta = W \cup b$ is the union of all the weights and biases of the NN. $MSE_{PDE} = \frac{1}{N_{PDE}} \sum_{i=1}^{N_{PDE}} |R_{PDE}(x_i, z_i, t_i)|^2$ is the loss term corresponding to the wave equation evaluated on a set of $N_{PDE}$ randomly chosen PDE training data $(x_i, z_i, t_i) \subset \Omega$ with $\Omega = \mathbb{R}^2 \times \mathbb{R}$ and

$$MSE_S = \frac{1}{N_{S_1}} \sum_{i=1}^{N_{S_1}} \left|R_{S_1}(x_i, z_i, t_i = t_1^0)\right|^2 + \frac{1}{N_{S_2}} \sum_{i=1}^{N_{S_2}} \left|R_{S_2}(x_i, z_i, t_i = t_2^0)\right|^2$$

represents the loss terms corresponding to the two vectorial early-time snapshot data $\vec{U_1^0}$ and $\vec{U_2^0}$ in terms of displacement. Similarly, the free-surface constraint and the observed data loss terms are defined as, $MSE_{P.C} = \frac{1}{N_{P.C}} \sum_{i=1}^{N_{P.C}} |R_{P.C}(x_i, t_i, z_i)|^2$ and $MSE_{Obs} = \frac{1}{N_{obs}} \sum_{i=1}^{N_{obs}} |R_{obs}(x_i, z_i, t_i)|^2$, respectively.

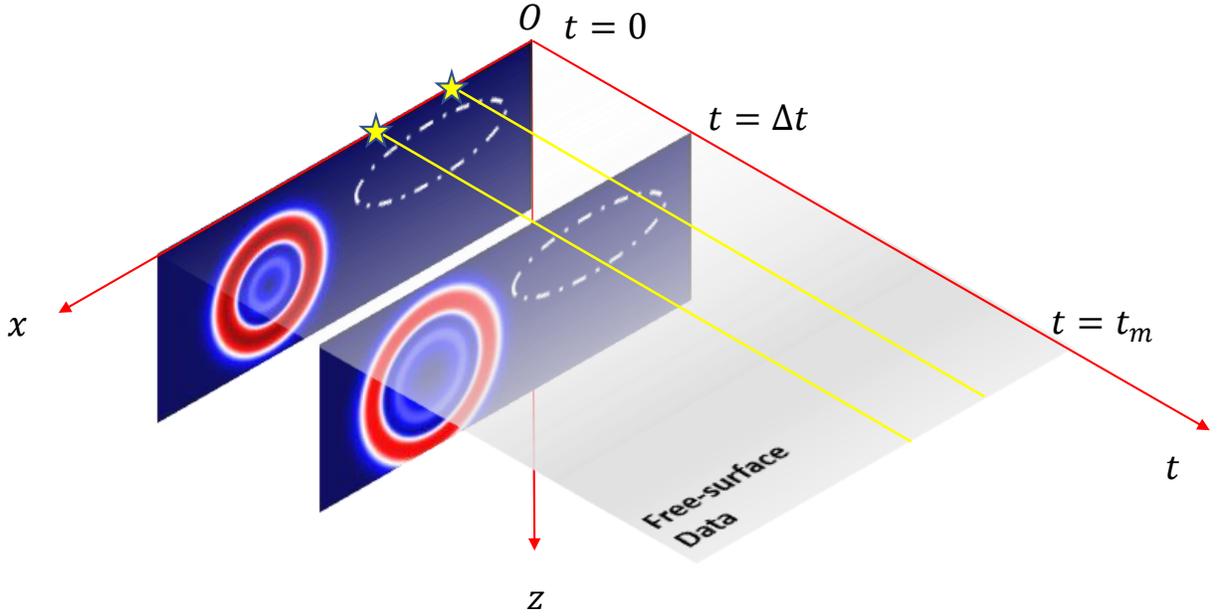

**Figure 2**. Schematic representation of a hypothetical computational domain (x, z and t) with PINN. The two time-snapshots at times 0 and $\Delta t$ are the only labeled data used from within the computational domain. The time-snapshots are color-coded for displacement amplitude. The white dashed line encloses a hypothetical heterogeneity. The grey hyperplane represents domain where the training data to apply, for example, a free-surface condition at the top of the domain. The two yellow stars represent the position of seismometers. $t_m$ is the duration of the time domain. The PDE training data is selected randomly from the entire computational domain. Note that for some of the case studies we only use a subset of the various data sets illustrated here.

In contrast with standard partial differential equations solvers applied to the wave equation, PINNs essentially cast the forward (and inverse) problems as an optimization with optimal weights and biases $\Theta^*$ obtained by minimizing equation 2, i.e., $\Theta^* = argmin\{MSE(\Theta)\}$.

The hyperparameters $\lambda_{i,s} > 0$ in equation 2 are set to normalize the different loss terms to guarantee a convergence to the correct solution. Failing to select proportionate loss terms would result in delayed convergence or possibly convergence to the wrong solution. Following the common practice in PINNs' literature, we find the proper values of $\lambda_{i,s}$ heuristically from trial and error; however, a more dynamic updating of these weights throughout the training process could be beneficial particularly for problems with persisting large residual errors at the boundaries (Wang et al., 2020). The learning rate annealing algorithm that Wang et al. proposed utilizes gradient statistics during the training process that would help maintaining a balance between different loss terms in equation 2.

**PINNs for the seismic inverse problem**
The goal of the inverse problem is to estimate the spatial distribution of the wavespeed $\alpha$ in equation 1 from the data collected at the surface or in a borehole by a set of seismometers. We use an additional NN with independent weights and biases to estimate the distribution of seismic wavespeed in the medium (Tartakovsky et al., 2020; Shukla et al., 2020; Haghighat et al., 2021). This secondary network is generally significantly smaller in size (depth and width of the network) than the primary network estimating the wavefield, since the structural complexity is much simpler in nature than the wavefield's variations in space and the distribution of wavespeed does not depend on time. We choose a fixed wavespeed NN architecture for all the inverse case studies in this paper, namely, a fully connected feed-forward NN with 5 hidden layers and 20 neurons per layer. An important outcome of the PINNs' formalism is that it eliminates the need for a user-defined parameterization of the computational domain for the inverse problem and the related biases imposed from such parametrization. However, one must make sure that the chosen inverse NN has enough layers and sufficient width for estimating the target structural heterogeneity one wishes to resolve- assuming that the observed data is not the limiting factor for the quality of the inversion.

It should be noted that enforcing both the free-surface constraint and the observed data from seismometers in the loss function (equation 2) might seem redundant at first glance. Although the data collected by the seismometers implicitly impose a free-surface condition, our numerical simulations show that it is necessary to include a loss term for the free-surface constraint for an accurate recovery of the reflected phases at the free surface. The main reason is that in practical scenarios, seismometers are limited to only sparse spatial coordinates on the surface while information from the free-surface constraint is readily available through a differential equation, i.e., $\nabla^2 \phi(x, t, z = 0) = 0$ that can be evaluated at any position on the free surface.

**Normalization**
There are two important normalization steps in PINNs' implementation that are crucial to guarantee a convergence to the correct solution. First, the input and output variables of the network must be mapped to the interval $[-1,1] \in \mathbb{R}$. Second, the acoustic wave PDE must be scaled such that both terms in equation 1 are on the same order. See appendix 1 for the appropriate formulations.

**Optimization**
Optimal values for the weights and biases of the proposed PINNs are obtained with the Adam optimizer (Kingma and Ba, 2014), an enhanced variant of the stochastic gradient descent with a learning rate of 1e-4 and a suitably chosen batch size (BSGD) for each case study. We stop the gradient descent search when the improvement from the previous epoch (iteration) becomes negligible, or when we surpass a fixed number of iterations. We have tested our results with L-BFGS (Liu and Nocedal, 1989), a second-order optimizer that takes advantage of the Hessian matrix but did not observe significant improvement in the training.

Table 1 summarizes the batch size for each of the specific loss terms from equation (2) for each case study. Note that unlike the PDE and the free-surface condition, we choose a fixed training data set to enforce the two early-time snapshots and the observed data from the seismograms. This is because the variability in the two latter data types is small enough to be captured with a relatively small data size. The optimal density of the different training data sets depends strongly on the ratio of the wavelength of the propagating wave to the domain size and the length scale of the structural heterogeneities as well as the total computational time.

**Validation**

We use numerical solutions obtained from SpecFem2D as the ground truth for the validation of the forward PINN model. We also use SpecFem2D to generate the early-time snapshot data and synthetic seismograms, so that we can test the PINN efficiency and accuracy. Any time-snapshots of the wavefield generated either analytically or with any numerical solver can be utilized and ultimately (in a future study) real data will replace the synthetically generated seismograms for the inversions. The SpecFem2D models are discretized on a 100x100 mesh spatially and we used a second-order explicit Newmark time stepping scheme with the time step-size 4e-5 seconds. We employed perfectly matched layers (PML) (Komatitsch and Tromp, 2003) at the boundaries of the domain with a thickness of 10 nodes to simulate the absorbing boundary conditions for the point source cases, and the Stacey absorbing boundary conditions for the teleseismic plane wave source cases (Stacey, 1988).

## Computational Experiments

In this section we examine the efficiency and accuracy of the PINN approach with the loss function given in (2). We present different case studies for forward and inverse modeling with the acoustic equation. We start with the forward simulation and then proceed to the inverse problems in the next section.

**Case 1. Application of PINNs to wave propagation in a heterogenous medium**

Since our main focus in this paper is FWIs, we restrict our application of PINN forward acoustic wave propagation to a single case of heterogeneous medium. Fig. 3 shows the wavespeed distribution in the domain with a background of $3 \, km/s$ perturbed with Gaussian heterogeneity of a minimum velocity of $2 \, km/s$ and a width (standard deviation) of 2.5 km. We choose a network with 4 hidden layers and 50 neurons per layer. The size of the training data set for the PDE loss term is 20,000 points while 3600 points are set for each of the early-time snapshot data, picked from a normal distribution in space and time. We note that, for this particular case study, we do not impose a free-surface condition at the top of the domain, since our objective here is to observe how well PINNs can capture the wave propagation and interaction with the prescribed heterogeneity in the forward problem. The weights for different loss terms in equation (2) are $\lambda_1 = 0.1, \lambda_2 = 1, \lambda_3 = \lambda_4 = 0$. We simulate 5 seconds of wave propagation. Fig. 4 shows the evolution of the history of convergence of loss terms in equation (2).

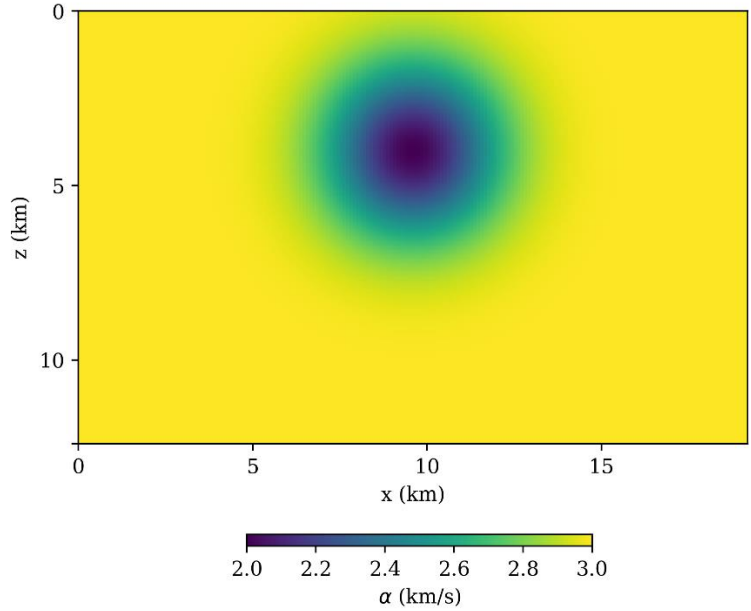

**Figure 3**. The distribution of the input wavespeed $\alpha$ for the forward problem

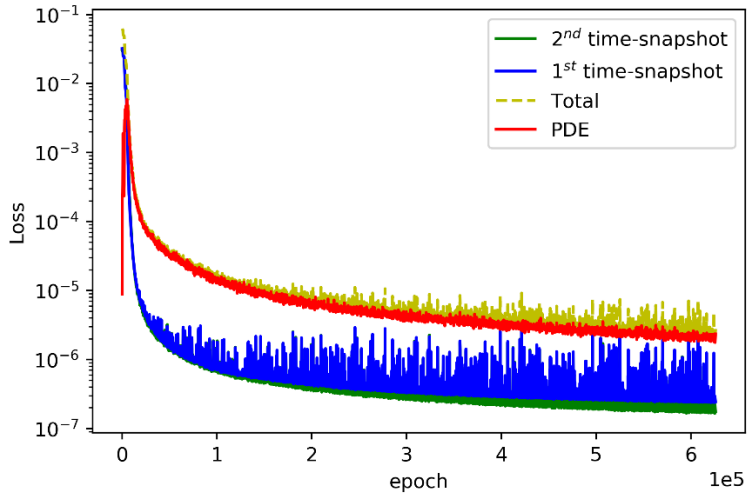

**Figure 4**. Evolution of the different loss terms in equation (2) as well as the total loss for the forward problem

Fig. 5 compares snapshots of the wavefield comparing PINN's prediction to the ground truth simulation using SpecFem2D. It is important to note that we are not using any labeled training data other than the two early-time snapshots of the wave propagation in the homogeneous regions of the domain, which serve to provide information about the acoustic source. In other words, the PDE is the sole information that the PINN uses to simulate the wave propagation in the heterogenous part of the domain. Fig. 5 shows that PINN provides a good solution to the wavefield. To avoid biases in network's parameters, all misfits are computed on a different input data set than the ones used for training.

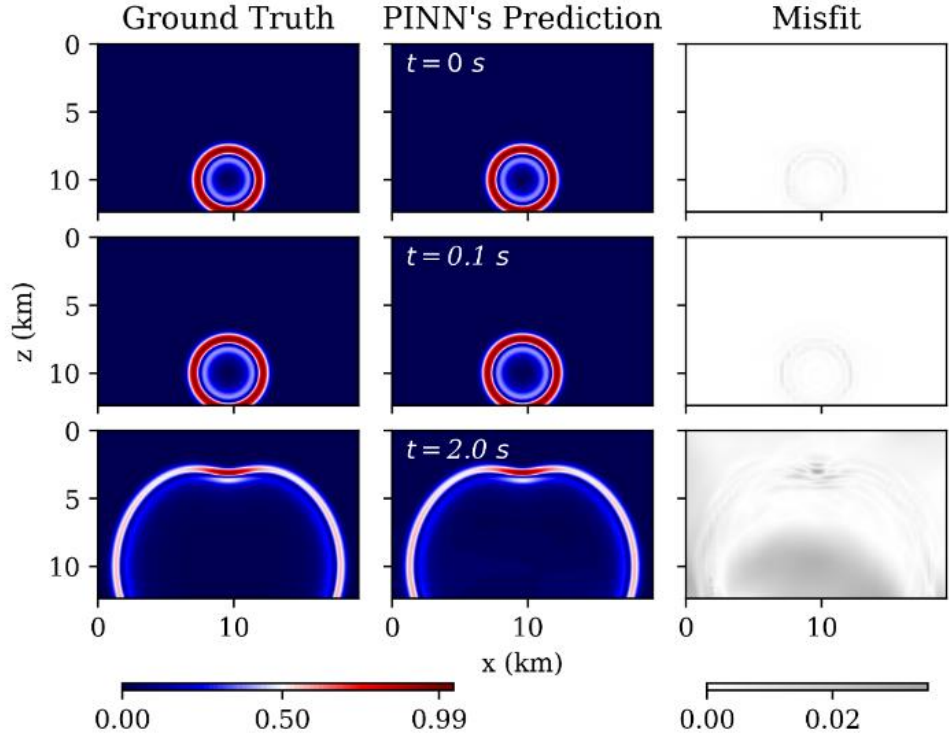

**Figure 5**. Ground truth versus predicted magnitude of the wavefields from PINN for the forward problem. The first two wavefield snapshots from SpecFem2D at $t = 0, t = 0.1\ s$ are used as the training data. The misfits show the absolute pointwise error.

**Application of PINNs to inverse modeling**

In this section, we present the application of PINNs to FWIs. For all case scenarios, we follow the same strategy outlined here. We build a 2-D domain with prescribed wavespeed distribution and simulate the acoustic wave propagation using SpecFem2D. We then use the generated seismograms as the data (observation) to run the inversion. Again, we use two early-time snapshots of the displacement field from SpecFem2D as the training data. It is important to note that we only use snapshots taken before the wave interacts with any heterogeneities in the ground truth model, so as to avoid providing more information into PINN than would normally be available in practical inverse problems encountered in seismology. In all but the last example, we impose a free-surface condition at the top of the domain. We test the PINN's performance on recovering discontinuous as well as continuous velocity anomalies with various seismic source types such as a single point source, collection of point sources and teleseismic plane waves. We also assess the PINN's ability to capture free-surface reflections and absorbing boundary conditions in these case studies.

**Case 2. Crosswell experiment: Homogeneous velocity model with a single point source**

We start with the simplest case study, a homogeneous domain with wavespeed $3\ km/s$. We use PINNs to retrieve the wavespeed across the domain in a synthetic crosswell experiment given the set of training data outlined above. We choose a network with 4 hidden layers and 50 neurons in each. The size of the training data set for the PDE loss is 10,000, each of the early wavefield time-snapshots adds 3600 data points, and the free-surface constraint finally adds another 5000, all chosen from normal distributions in space and time. We use data points from a 0.4 second time window with a sampling frequency of 200 Hz from each component of the seismograms' time series and from 20 equally distanced seismometers (total 3240 data points from all seismometers) with depths ranging from the top of the model and extending to 450 m (Fig. 7a). The seismic wavefield is generated from one point-source with a Gaussian source time function and a dominant frequency of 20 Hz located on the left side of the domain (Fig. 8). The weight values for different loss terms in equation 2 are $\lambda_1 = \lambda_3 = 0.1, \lambda_2 = \lambda_4 = 1$. Fig. 6 shows the evolution of the different loss terms until convergence.

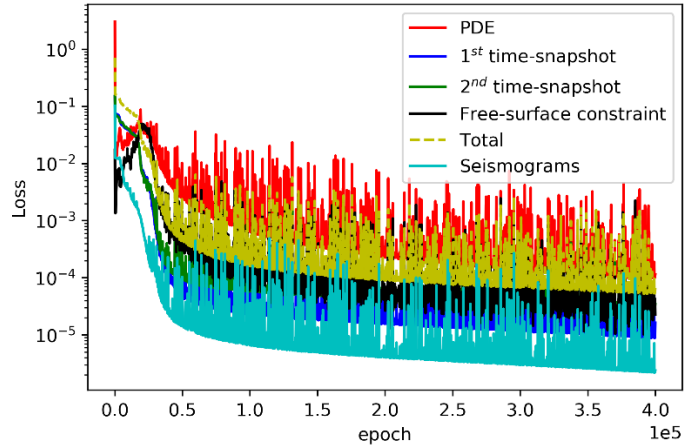

**Figure 6**. Evolution of the different loss terms in equation 2 as well as the total loss for the synthetic crosswell experiment in a homogeneous domain.

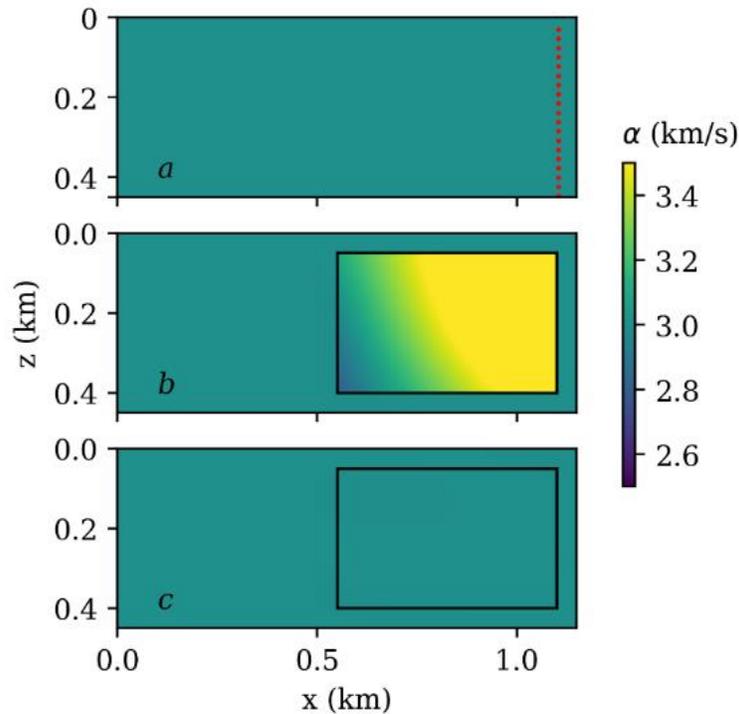

**Figure 7.** (a) True domain, (b) Initial guess (c) PINN's inversion results after convergence. The black rectangles in (b) and (c) show the area where the inversion is performed with PINN. The red dots in (a) show the locations of the seismometers.

Despite the relatively poor initial guess for the wavespeed, we can see that PINN successfully recovers an accurate estimate of the domain's wavespeed (Fig. 7) as well as the wavefield (Fig. 8). The quality of the solution is further supported by the great match between the synthetic (observation) and calculated seismograms (Fig. 9).

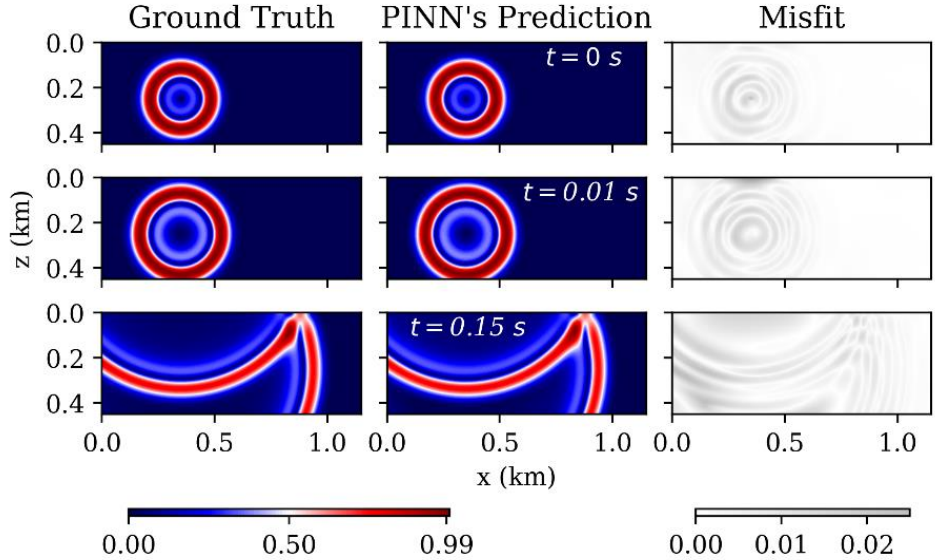

**Figure 8.** Comparison between ground truth and modeled wavefields and their absolute pointwise differences for the synthetic crosswell experiment with a homogeneous wavespeed distribution. The first two wavefield snapshots from SpecFem2D at $t = 0, t = 0.01\ s$ are used as training data.

At the top surface, where we imposed free-surface constraints, we observe a good match between the reflected wave simulated with SpecFem2D applied to the true wavespeed model and the PINN solution in terms of amplitude, waveform, and timing (Fig.8).

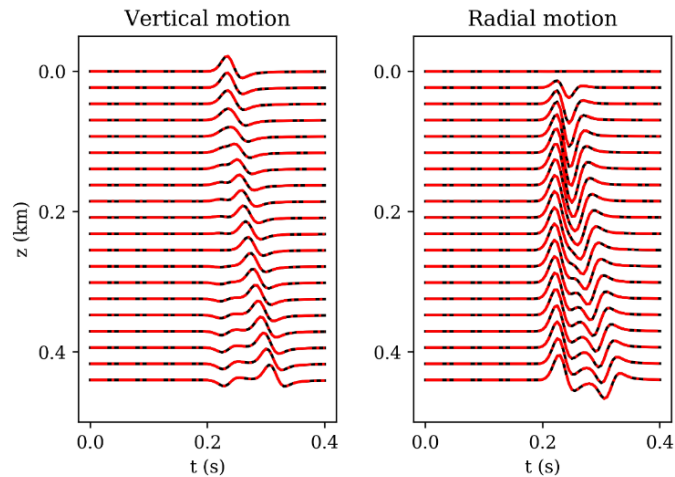

**Figure 9.** The comparison between ground truth (black line) and PINN's prediction (red dashed line) of the vertical and radial component seismograms for the synthetic crosswell experiment with a homogeneous wavespeed distribution. Locations of the seismometers are shown in fig. 7a.

It is important to emphasize that we have not explicitly enforced absorbing conditions in the loss function. The ability of PINN to manage automatically absorbing conditions stem from PINNs solutions being smooth and infinitely differentiable functions. Both components of the time series of the synthetic seismograms (observations) are also matched accurately by the PINNs solution (Fig. 9).

**Case 3. Crosswell experiment: Ellipsoidal velocity anomaly with a single point source**

This case study is designed to test the ability of PINNs to recover sharp 2-D anomalies from a synthetic crosswell experiment. An ellipsoidal low velocity anomaly with wavespeed of $2\ km/s$ is embedded in a homogeneous background model ($3\ km/s$). The velocity contrast between the anomaly and background is sharp (step function). The size of the domain, the location of the source and seismometers and the sampling frequency of the training data are the same as the previous case (Fig. 10a). We design a network with 8 hidden layers and 100 neurons per layer for the wavefield NN. The size of the training data set for the PDE loss is increased to 40,000 points and remains 3600 and 5000 for each of the early-time snapshot data and the free-surface constraint, respectively, again chosen from a normal distribution in space and time. Furthermore, we set the weights of the misfit function $\lambda_1 = \lambda_3 = 0.1, \lambda_2 = \lambda_4 = 1$.

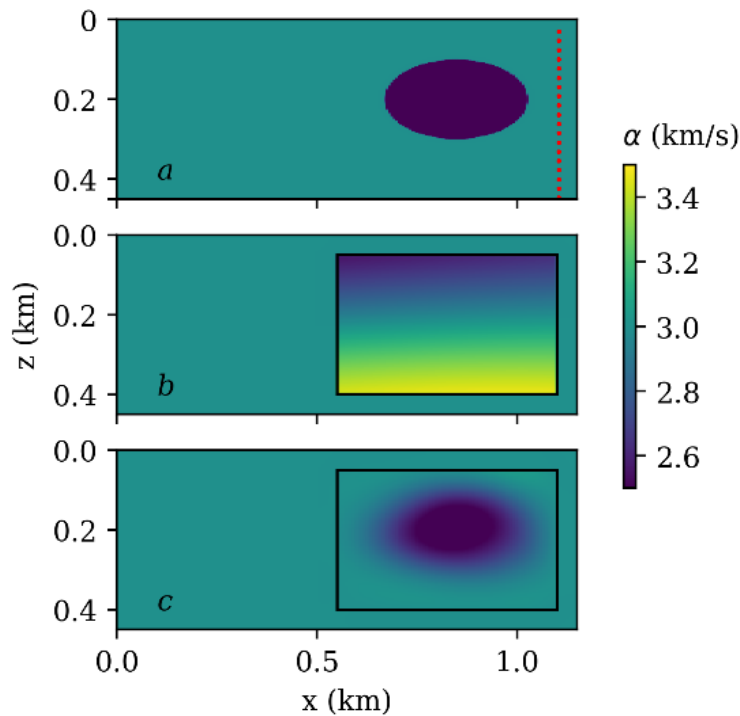

**Figure 10**. (a) True domain, (b) Initial guess (c) PINN's inversion results after convergence. The black rectangles in (b) and (c) show the area where the inversion is performed with PINN. The red dots in (a) show the locations of the seismometers. Note in the true domain the wavespeed transition from the ellipsoidal anomaly to the background is discontinuous.

Fig. 10 shows the inverted solution for the acoustic wavespeed in comparison to the ground truth and the starting model. PINN successfully retrieves the location, dimension, and magnitude of the anomaly. The inverted solution is smoothed instead of the sharp discontinuous transition in material property of the true model.

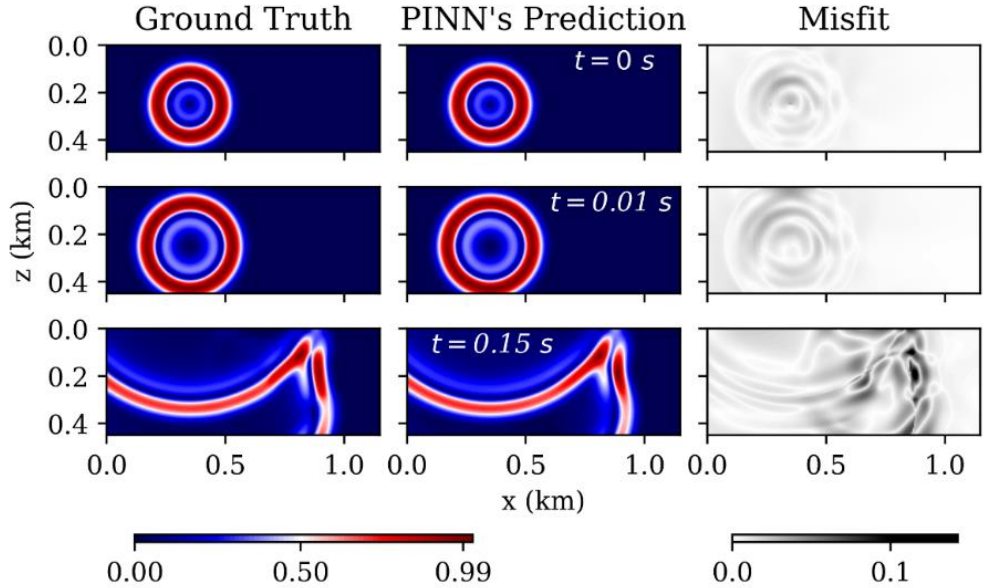

**Figure 11**. Comparison between ground truth and modeled wavefields and their absolute pointwise differences for the synthetic crosswell experiment with a discontinuous ellipsoidal anomaly. The first two wavefield snapshots from SpecFem2D at $t = 0, t = 0.01\ s$ are used as the training data.

A comparison of the wavefield after 0.15 seconds is provided in fig. 11. It shows that the inverted solution matches the wavefield of the synthetic true solution accurately. The match between observed and modeled seismograms for each of the 20 seismometers is also excellent (Fig. 12). For some of the seismograms, the slightly larger discrepancies between the input data and the outputs from PINN around the end of the time window analyzed is a regression artifact independent of the method. There, the optimization close to the final time is not perfect. Nevertheless, the prominent parts of the time series have been closely matched by PINNs solution and increasing the sampling frequency would improve the remaining misfit.

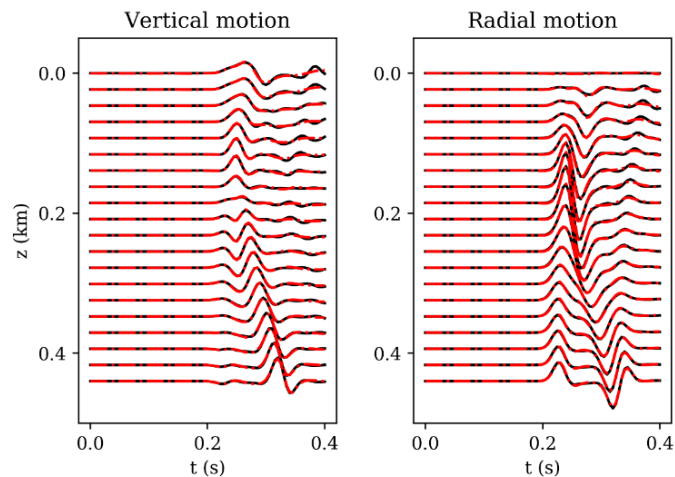

**Figure 12**. The comparison between ground truth (black line) and PINN's prediction (red dashed line) of the vertical and radial component seismograms for the synthetic crosswell experiment with a discontinuous ellipsoidal anomaly. Locations of the seismometers are shown in fig. 10a.

We have also performed this inversion with a smaller NN with 4 hidden layers and 50 neurons per layer. The smaller network still yields a good estimate of the wavespeed distribution, however the recovered wavefield is less accurate and justifies the choice of a deeper network for an improved accuracy (See Supporting Information).

**Case 4. Ellipsoidal anomaly with two incident plane waves**
In teleseismic imaging, plane waves from multiple incidence angles are generally required for a suitable ray coverage of the area under study. In the next example, we demonstrate how to incorporate multiple events into PINNs' formulation in a consistent manner such that it avoids substantial increase in computational costs. Exploiting the linearity of the acoustic wave equations, we superpose multiple events and design a network to simulate the resulting superposition of all events at once. This approach avoids defining a network for each event, which can become prohibitively expensive.

In this case study, we generate two compressional plane waves with a ricker source time function and a dominant frequency of 2 Hz with incidence angles + 20 and -20 degrees (with respect to the vertical axis). The two plane waves propagate through an ellipsoidal low velocity anomaly (wavespeed of $2\ km/s$) embedded in a homogeneous background model with $3\ km/s$. We set up 17 equally spaced seismometers at the top surface of the model starting from the top leftmost corner of the domain and extending 15 km to the right (Fig. 13a). We record 5 seconds of the seismic signal with a sampling frequency of 20 Hz for training the PINN (3400 training data from the seismometers' time series). A PINN with 8 hidden layers and 100 neurons per layer is used for the wavefield neural network. The training data sets for the loss terms corresponding to PDE, free-surface constraint on top and each of the wavefield early time-snapshots consist of 40,000, 5000 and 3600 data points, chosen from a normal distribution in space and time, respectively. Furthermore, we set the weights of the misfit function $\lambda_1 = 0.01, \lambda_2 = \lambda_3 = \lambda_4 = 1$. Fig. 13 demonstrates the excellent agreement in terms of location, dimensions, and magnitude of the anomaly between the ground truth and PINN's inversion results.

The amplitude and structure of the recovered wavefield from PINN follows closely the forward simulations obtained from the numerical solver (Fig. 14). However, we observe a slight timing difference at different parts of the wavefield at later times (Fig 14, lower most right panel). This time-difference is due to the fact that PINN recovers a smooth version of the structural heterogeneity (as in any other inverse technique), and since it solves the forward and inverse problems in a coupled fashion, the forward solution (wavefield) takes effect from the smoothed inverted structure. The smoothening associated with the inversion is also responsible for the absence of certain phases reflected off the anomaly. On the other hand, PINN is able to capture the reflected waves off the free surface on top (Fig. 14). This shows that when the interface condition is explicitly enforced, PINN can capture reflections. Moreover, we observe the PINNs' great ability in simulating absorbing boundary layers at the left, right and bottom edges of the computational domain, without explicitly enforcing them (Fig. 14). Our results show that the bulk of the training is spent on improving small discrepancies, while PINN converges to a satisfying velocity model in less than 70,000 epochs, instead of the 400,000 used for a complete training in this case (See supporting information).

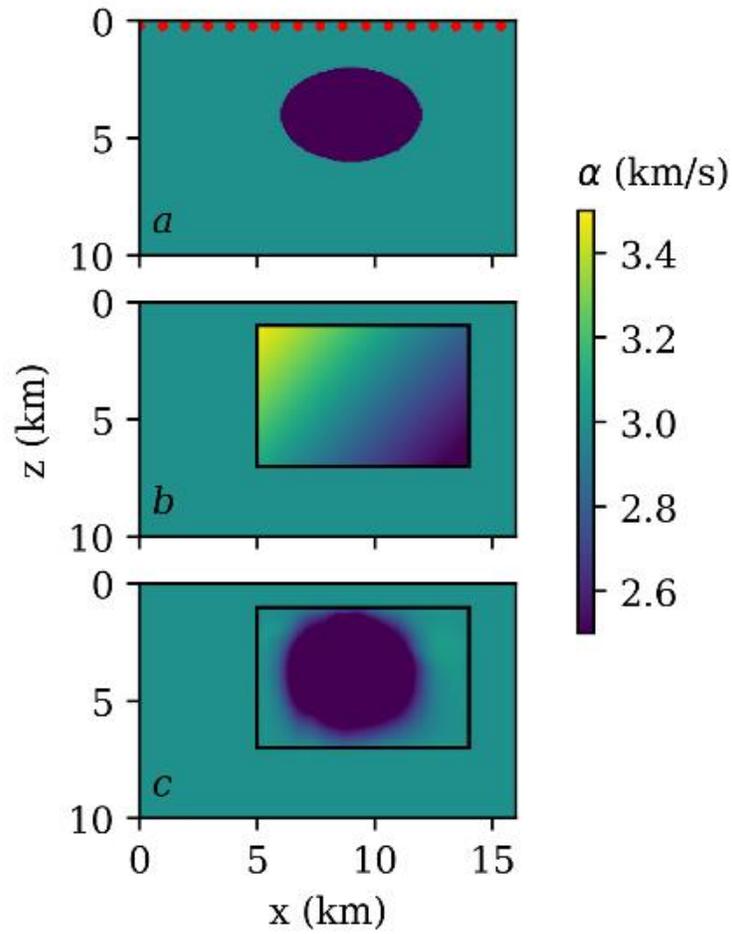

**Figure 13.** (a) True domain, (b) Initial guess (c) PINN's inversion results after convergence for the teleseismic case study. The black rectangles in (b) and (c) show the area where the inversion is performed with PINN. The red dots in (a) show the locations of the seismometers.

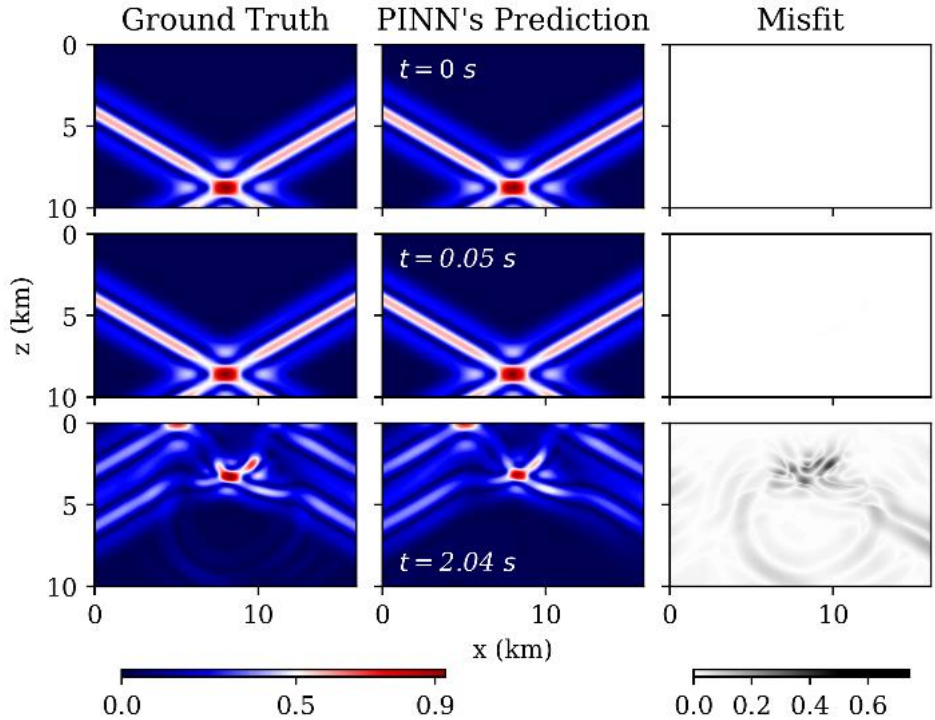

**Figure 14.** Comparison between ground truth and modeled wavefields and their absolute pointwise differences. The first two snapshots from SpecFem2D at $t = 0, t = 0.05\ s$ are used as the training data.

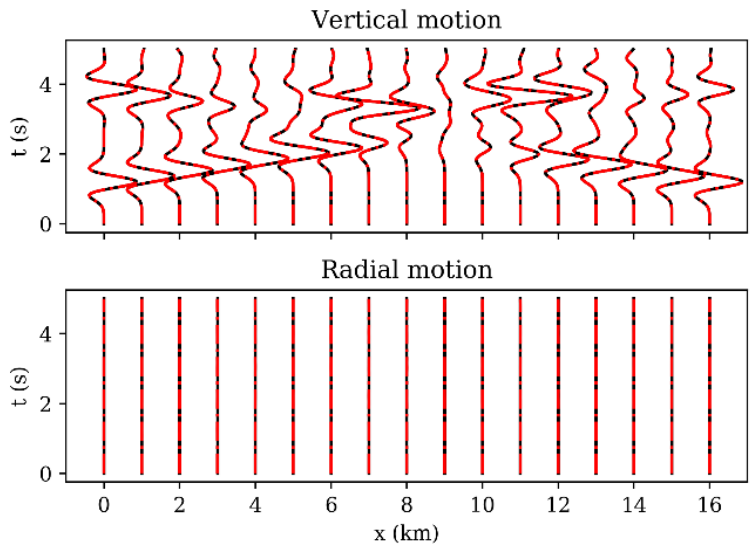

**Figure 15.** The comparison between ground truth (black line) and PINN's prediction (red dashed line) of the vertical and radial component seismograms for the teleseismic case study. Locations of the seismometers are shown in fig. 13a.

From Fig. 15, we observe that PINN finds a precise fit to the observed synthetic seismograms ensuring a correct capture of the physics of the wave's interaction with different parts of the domain. For instance, there is a sharp deficit in the recorded energy from the observed vertical component seismograms at around 9 km, signaling a "shadow" region affected by a low velocity anomaly at depth, which is impressively captured by PINNs as well. On the other hand, the absolute zero energy on the radial component seismograms, signature of a free-surface physical constraint in acoustic media is also correctly retrieved by PINN.

**Case 5. Checkerboard test with nine point-sources**

For the last case study, we test PINNs' performance when inverting for a sinusoidal checkerboard model velocity with alternating positive and negative anomalies, as depicted in Fig. 16a. We also aim to see if a reasonably sized NN can capture the complex response of the system to as many as nine point-sources. The point locations for 20 seismometers are set at the top surface at equal distance from each other, starting at 2.6 km from the left side of the domain extending to 17.6 km to the right. We use nine point-sources all with a Gaussian source time function and a dominant frequency of 2 Hz (Fig. 17). We use a NN with 10 hidden layers and 100 neurons per layer. The size of the training data sets for the PDE, and each of the early-time snapshots are 60,000 and 3600, respectively. We also use data from the seismometers with a sampling frequency of 50 Hz (11600 total data points from seismometers' time series). The different loss term weights are set $\lambda_1 = 0.1, \lambda_2 = \lambda_3 = 1, \lambda_4 = 0$. For this case we do not impose a free-surface constraint at the top of the domain. From fig. 16 and 17 we observe that PINN is capable of recovering a complex oscillatory velocity gradient and the corresponding wavefield. The synthetic seismograms from SpecFem2D have also been successfully matched by PINN's outputs with negligible misfit.

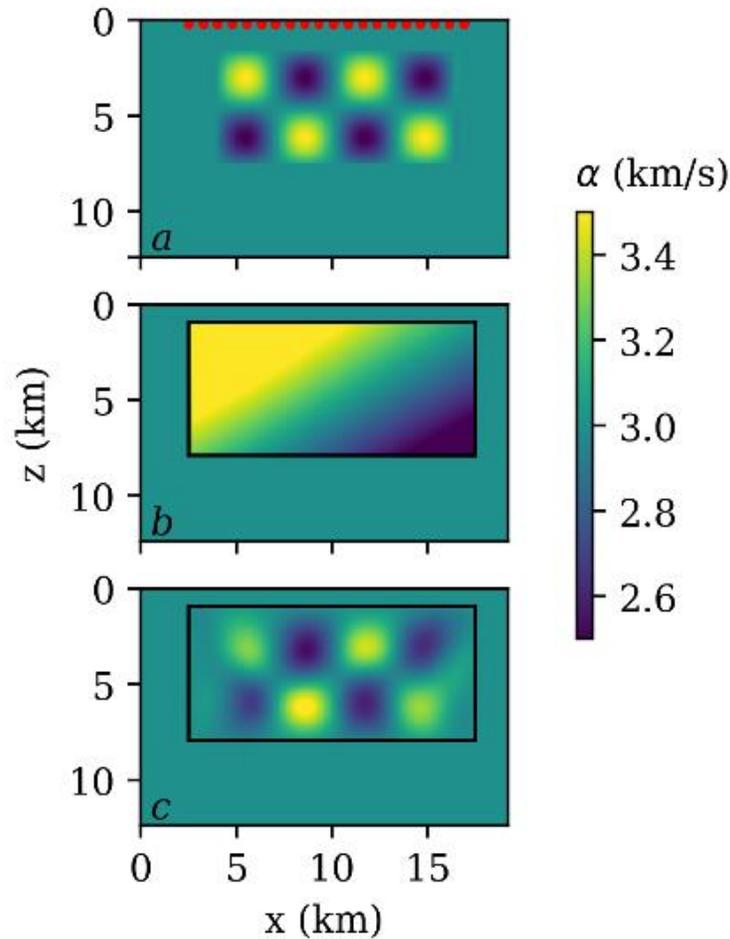

**Figure 16.** (a) True domain, (b) Initial guess (c) PINN's inversion results after convergence. The black rectangles in (b) and (c) show the area where the inversion is performed with PINN. The red dots in (a) show the locations of the seismometers.

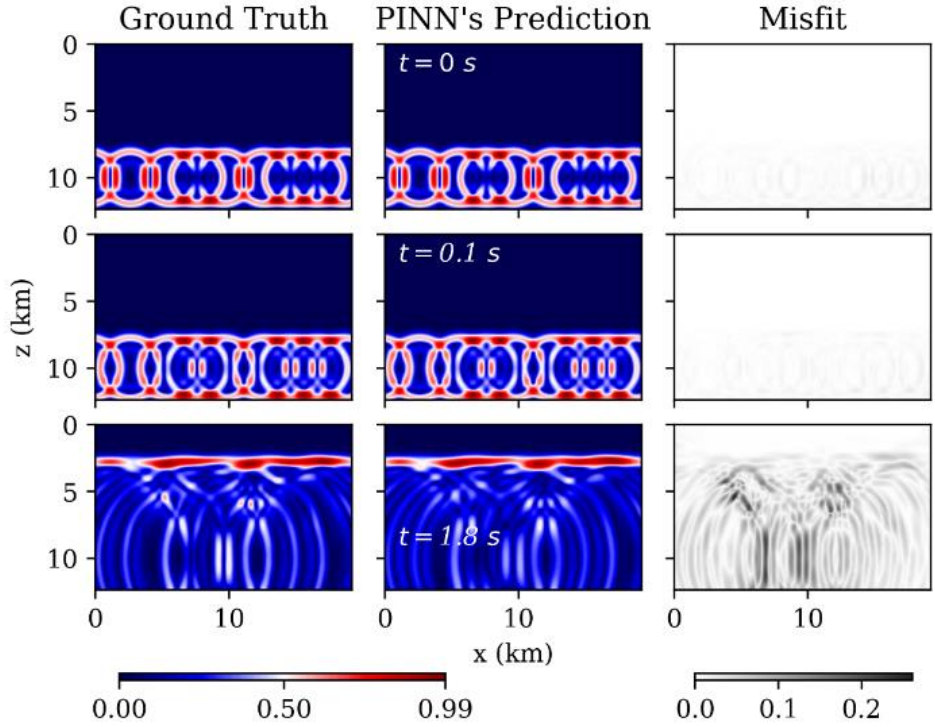

**Figure 17.** Comparison between ground truth and modeled wavefields and their absolute pointwise differences. The first two snapshots from SpecFem2D at $t = 0, t = 0.1\ s$ are used as the training data.

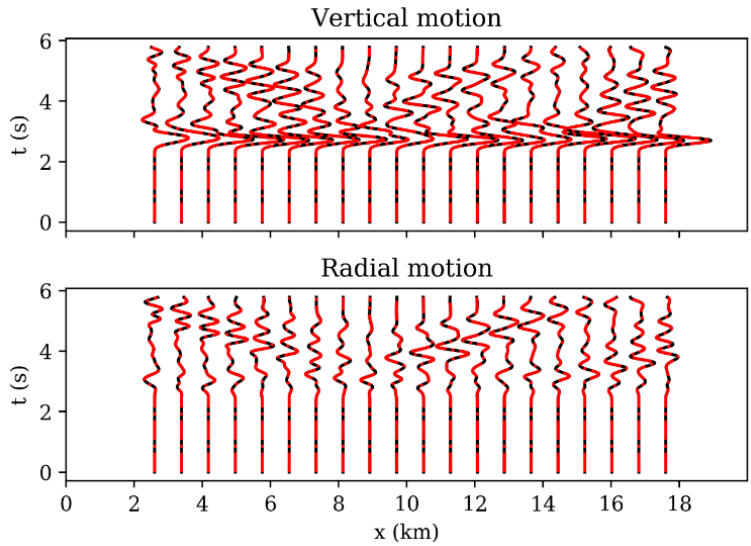

**Figure 18.** The comparison between ground truth (black line) and PINN's prediction (red dashed line) of the vertical and radial component seismograms for the checkerboard case study. Locations of the seismometers are shown in fig. 16a.

# Discussion

We have shown that PINNs are able to accurately solve the forward and inverse modeling of the acoustic wave equation in complex media. PINNs are meshless and have impressive generalization capability and given only a set of two early-time snapshots they can predict the wavefield solution much later in time. Also, with minimal observed data only at the surface, and without any training data within the computational domain (other than the early-time snapshots) PINNs yield excellent results for seismic inversions. This shows that the PINN formalism can be implemented independent of other numerical solvers that require more complex implementation.

We extended the original approach by Raissi et al., 2019 by incorporating multiple seismic events into one NN and hence greatly optimizing the inverse problem for practical applications. Different seismic sources (plane waves and point sources) with variable frequencies can be implemented with limited effort. However, it must be noted that if the energy from the superposition of seismic sources interferes destructively at the site of a seismometer, superposition must be avoided because of the loss of information and the consequent deterioration of the solution. In this case, a possible approach is to define different groups of non-destructively interfering events and then define an independent NN for each group. This approach will still decrease the computational costs compared to defining a NN for each event.

Our study shows that the computational costs (size of the NNs and number of epochs to reach convergence) of the problem is strongly influenced by the frequency of the source and the complexity of the wavespeed structures. We observe that a smaller NN is capable of yielding satisfactory results for an inverse problem if the structure of the medium is simple. For instance, the NN for the homogenous inversion model (Case 2) is significantly smaller than the NN for the model with an ellipsoidal anomaly (Case 3). The NN's setup is also influenced by the frequency content of the source. Modeling higher frequencies (for the same physical dimensions) requires a larger training data set for the PDE loss to capture finer features such as refracted waves which are more pronounced at higher frequencies.

PINNs are a meshless method. They offer great flexibility in terms of implementation and if *a priori* knowledge is available for inverse problems. Moreover, our results show that PINNs perform well even without *a priori* knowledge or in the absence of an educated starting model. Nevertheless, one can easily implement *a priori* constraints on the wavespeed distribution. For instance, if we expect a medium with dominantly vertical variations in acoustic/seismic velocities, it is straightforward to set the functional form of the inverse NN to explicitly depend only on one spatial variable, namely z. This can be achieved with no fundamental change to the original algorithm.

PINN's implementation is made simple and compact thanks to modern python's libraries such as TensorFlow and PyTorch. For example, the script (available on https://doi.org/10.26300/x3wd-4k56) utilizes only a few hundred lines to run both the wave propagation and optimization for the inversion. It is also easy to modify or change the constraints imposed at the boundaries of the domain as they are expressed only in the definition of one of the loss terms in equation (2).

For the case studies with a discontinuous true model, reducing the smoothening effect of the inversion is possible if one uses a significantly larger training data set for the PDE part of the loss function and includes *a priori* knowledge of the presence of a discontinuity or a supervised distribution of many training data around the discontinuity. Moreover, a different type of minimization norm, e.g. L1, may lead to sharper discontinuities.

A challenge when using PINN comes from the heuristics nature of the algorithm when selecting the proper weights for each loss terms or selecting the proper network size. Work remains necessary to better constrain the factors that control the convergence of PINNs and the criteria for choosing the minimum density of the training data for the PDE to guarantee convergence to the correct solution (Shin et al., 2020; Wang et al., 2020); see also the recent work of McClenny and Braga-Neto (2020). Moreover, for large computational domains, efficient strategies to implement PINNs on GPUs are required to deal with the large memory cost involved. The implementation of PINNs with a domain decomposition in space and space-time domain is successfully studied by Jagtap et al. (2020) and Jagtap and Karniadakis (2020), which is further extended to a multi-GPU platform by Shukla et al. (2021). Therefore, the computational efficiency of proposed method for larger and elastic approximation can be successfully tamed .

**Conclusion**

We have demonstrated that physics-informed neural networks are able to solve the wave propagation and full waveform inversions by relying on information from governing partial differential laws when there is limited data available. We have shown how inversions with multiple events can be performed with PINNs with limited additional memory or computational cost. PINNs' seamless ability to handle diverse sets of constraints, their meshless nature and the simplicity in formulation/implementation open new perspectives as the next generation of inverse solvers for geophysical applications. They provide a flexible framework to incorporate multiple data types and/or any apriori knowledge of the structure imaged and hence offer a great potential for joint and Bayesian inversions (Yang et al., 2021).

| Type | Inhomogeneity type | Source type & Frequency | Network [Width]*Depth | Activation function | $N_{pde}$ | $N_S$ | $N_{P.C}$ | $N_{Seismograms}$ |
|---|---|---|---|---|---|---|---|---|
| **Forward** | Gaussian Anomaly | 1 Point source 2Hz | [50]*4 | Sin() | 20,000 | 2 *3600 | N/A | N/A |
| **Inverse** | Homogeneous | 1 Point source 20Hz | [50]*4 | tanh() | 10,000 | 2 *3600 | 5000 | 2*1620 |
| **Inverse** | Ellipsoidal Anomaly | 1 Point source 20Hz | [100]*8 | tanh() | 40,000 | 2 *3600 | 5000 | 2*1620 |
| **Inverse** | Ellipsoidal Anomaly | 2 Plane waves 2Hz | [100]*8 | tanh() | 60,000 | 2 *3600 | 5000 | 2*1700 |
| **Inverse** | Sinusoidal Checkerboard | 9 Point sources 2Hz | [100]*10 | Sin() | 60,000 | 2 *3600 | N/A | 2*5800 |

**Table 1**: Parameters used for the simulations. [50]*4 indicates 4 hidden layers and 50 neurons in each. Note that the NNs in this table indicate the forward NN. We use the same wavespeed NN for all the case studies here with 4 layers and 20 neurons in each. $N_{Seismograms}$ is the total number of training data from all the input seismograms in both x and z directions. All the input data sizes denote the batch size for the corresponding loss term.

**Appendix A**

Here we show the two important normalizations used in PINN's formulation in this study:

a) We map each input variable to the neural network onto the interval $[-1,1] \in \mathbb{R}$ (Raissi et al., 2019):

$$X \rightarrow \frac{2X}{\max(X)} - 1$$

b) We scale the wave PDE in equation (1) to constrain the wavespeed in the $[0,1] \in \mathbb{R}$ interval. Using the scaled spatial coordinates $x = \max(\alpha)\, x^\star$, $z = \max(\alpha)\, z^\star$, yields,

$$\alpha^{\star 2} \nabla^{\star 2} \phi = \frac{\partial^2 \phi}{\partial t^2}$$

where $\alpha^\star = \alpha/\max(\alpha)$ and $\nabla^{\star 2} \equiv \frac{\partial^2}{\partial x^{\star 2}} + \frac{\partial^2}{\partial z^{\star 2}}$.

Acknowledgment: We thank Victor C. Tsai for the helpful discussions on the set-ups of the synthetic case studies. This research was conducted using computational resources and services at the Center for Computation and Visualization, Brown University. A sample script in python for this study can be found on Brown University's data repository: https://doi.org/10.26300/x3wd-4k56.

**Supporting information**

In this section we discuss two main points regarding training PINNs for FWIs: (a) How does using a smaller NN affect the inversion results? (b) How accurate is a PINN that has been trained for a smaller number of epochs for the same number of hidden layers and neurons? We perform two additional inversions to address each of these questions.

To address the first question, we design a smaller NN with 4 hidden layers and 50 neurons in each layer to redo the inversion for the case study with an ellipsoidal anomaly and a 20 Hz point source (Case 2. In the main text). All parameters of the system and PINN's setup are the same as in Case 2. From fig. S1 we can see that the main features of the anomaly (location, approximate size, and strength) have been recovered well by PINNs, despite an uneducated initial guess. However, the smearing at the boundaries of the anomaly is larger than Case 2 in the main text with the larger NN. From fig. S2 we can also see that the estimation error for the wavefield has slightly increased compared to Case 2 in areas close to the boundaries of the velocity anomaly. The fit to the seismograms has not been affected greatly, which is confirmed from fig. S3. This exercise shows that it is important for the chosen NN to be expressive enough (large enough number of layers and neurons) for a correct estimation of the ground truth wavespeed and the wavefield.

To understand the effect of a shorter training episode, we record the results of the inversion for the ellipsoidal anomaly with two teleseismic plane waves (Case 4 from the main text) after 70,000 epochs, instead of 400,000. Fig. S4 shows that even with a shorter training process, the inverted wavespeed is acceptable, however with slightly larger smearing at the boundaries of the velocity anomaly compared to fig. 13c in the main text. Furthermore, fig. S5 shows that the overall shape of the waveforms is well preserved, however with slightly bent wavefronts (instead of straight plane waves) at later times. From fig. S5 we can also see that the free-surface constraint and hence the resulting reflection is well captured by a shorter training process. Additionally, there is no noticeable violation of the absorbing boundary conditions at the right, left and bottom boundaries. We can also see that the seismograms are fit equally well for the shorter training of PINNs (Fig. S6).

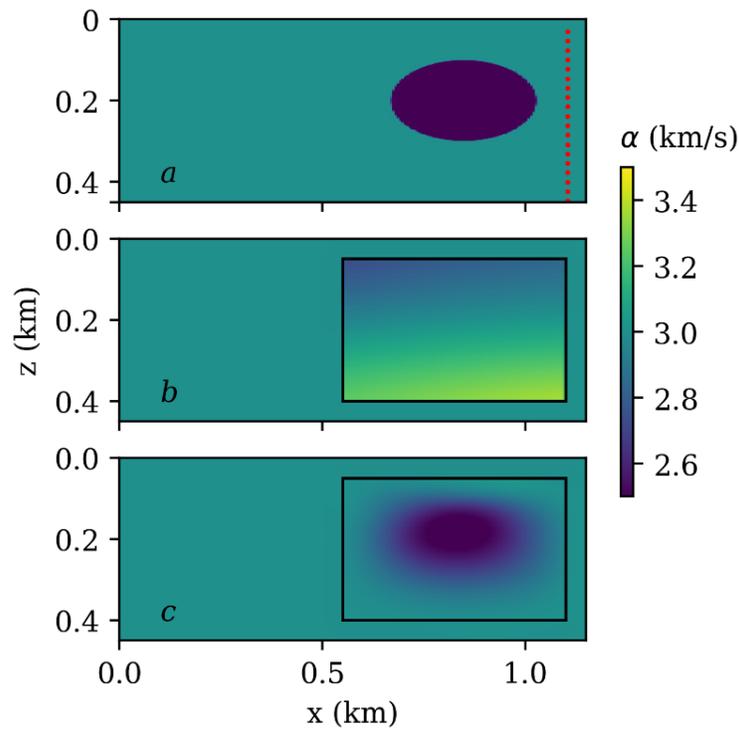

**Figure S1.** (a) Ground truth, (b) Initial guess (c) Inversion results for the synthetic crosswell experiment. A PINN with 4 hidden layers and 50 neurons in each layer has been used. The black rectangles in (b) and (c) show the area that we have inverted for with PINN. The red dots in (a) show the locations of the seismometers.

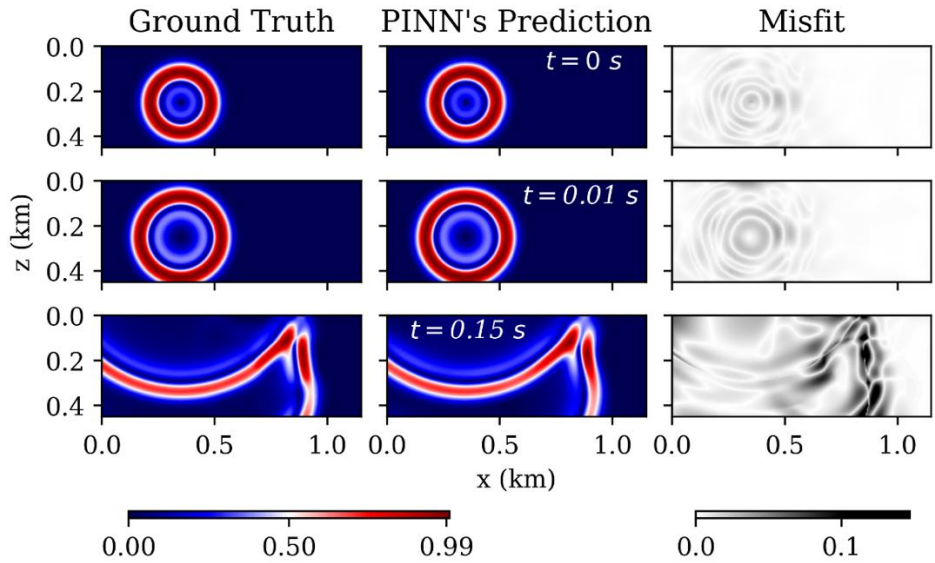

**Figure S2**. Wavefields for the synthetic crosswell experiment. Ground truth versus predicted magnitude of the wavefields from PINN and their absolute pointwise differences with a NN with 4 hidden layers and 50 neurons in each layer.

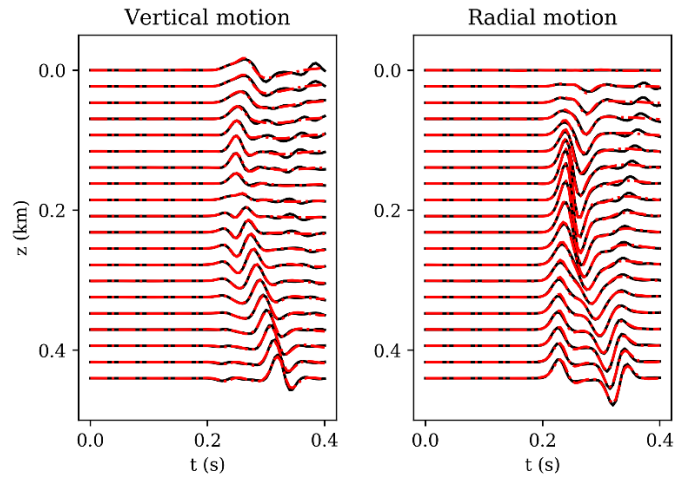

**Figure S3**. Vertical (left) and radial (right) motion seismograms for the synthetic crosswell experiment. black line: input, red dashed line: PINNs' prediction.

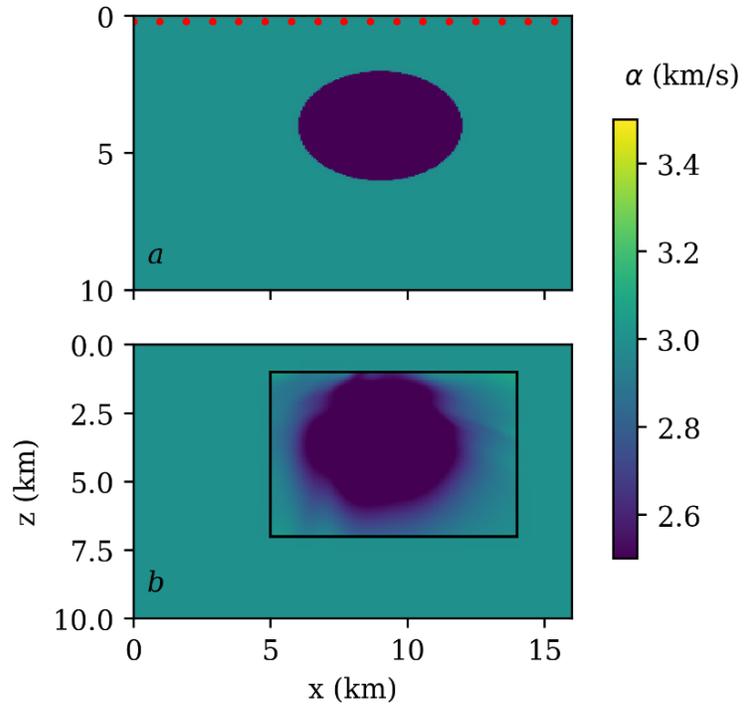

**Figure S4**. (a) Ground truth (b) inverted acoustic wave velocity from PINN after 70,000 epochs of training, for the teleseismic case study.

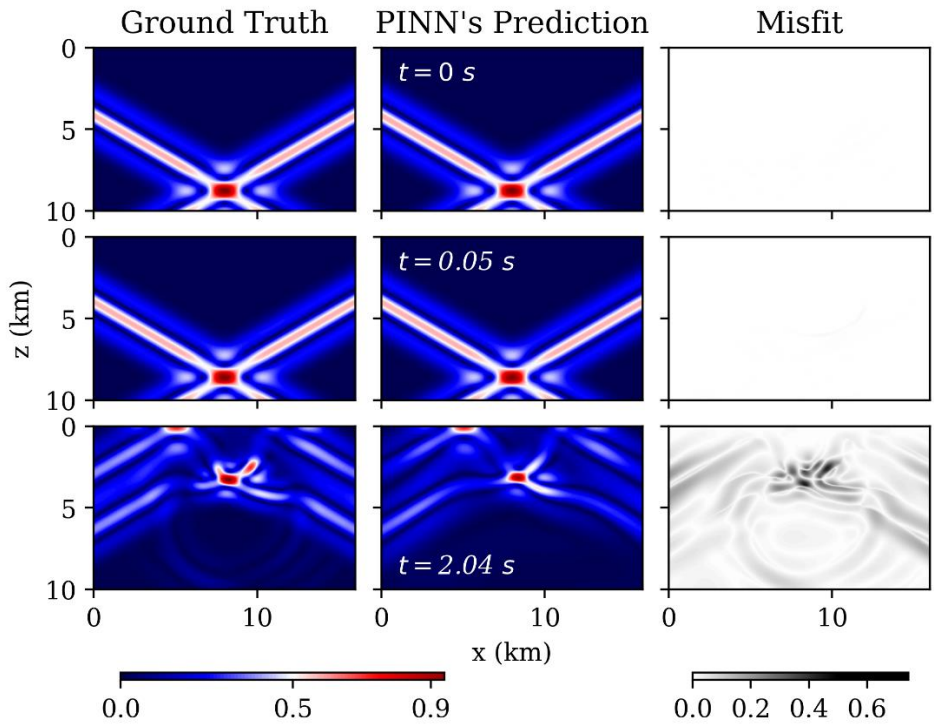

**Figure S5**. Ground truth versus predicted magnitude of the wavefields from PINN and their absolute pointwise differences after 70,000 epochs.

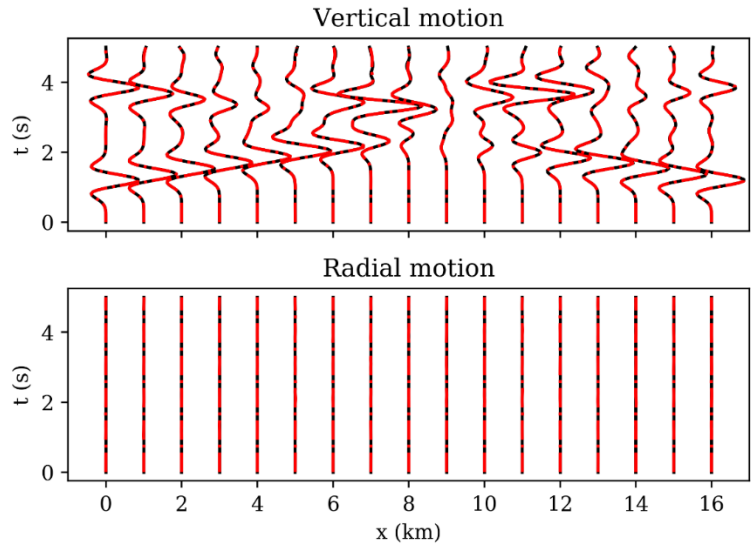

**Figure S6**. Vertical (top) and radial (bottom) motion seismograms after 70,000 epochs, for the teleseismic case study. black line: input, red dashed line: PINNs' prediction.